\documentclass[journal,10pt]{IEEEtran}

\usepackage{graphicx}
\usepackage{subcaption}
\usepackage{amsmath,amsfonts}
\usepackage{algorithmic}
\usepackage{algorithm}
\usepackage{array}
\usepackage{textcomp}
\usepackage{stfloats}
\usepackage{url}
\usepackage{verbatim}
\usepackage{booktabs}
\usepackage{tikz}
\usetikzlibrary{automata, positioning, fit, decorations.text}
\usepackage{textcase} 
\usepackage{float}
\usepackage{multirow}
\usepackage{makecell}
\usepackage{tabularx}
\usepackage{cite}
\usepackage{etoolbox}  

\usepackage[hidelinks]{hyperref}  

\begin{document}

\title{Factorized Deep Q-Network for Cooperative Multi-Agent Reinforcement Learning in Victim Tagging}

\author{Maria Ana Cardei and Afsaneh Doryab
\thanks{Maria Ana Cardei and Afsaneh Doryab are with the Department of Computer Science, University of Virginia, VA, 22904, USA. Correspondance: cbr8ru@virginia.edu, ad4ks@virginia.edu}
}



\maketitle

\begin{abstract}
Mass casualty incidents (MCIs) are a growing concern, characterized by complexity and uncertainty that demand adaptive decision-making strategies. The victim tagging step in the emergency medical response must be completed quickly and is crucial for providing information to guide subsequent time-constrained response actions.
In this paper, we present a mathematical formulation of multi-agent victim tagging to minimize the time it takes for responders to tag all victims. Five distributed heuristics are formulated and evaluated with simulation experiments. The heuristics considered are on-the go, practical solutions that represent varying levels of situational uncertainty in the form of global or local communication capabilities, showcasing practical constraints. We further investigate the performance of a multi-agent reinforcement learning (MARL) strategy, factorized deep Q-network (FDQN), to minimize victim tagging time as compared to baseline heuristics.
Extensive simulations demonstrate that between the heuristics, methods with local communication are more efficient for adaptive victim tagging, specifically choosing the nearest victim with the option to replan. Analyzing all experiments, we find that our FDQN approach outperforms heuristics in smaller-scale scenarios, while heuristics excel in more complex scenarios. Our experiments contain diverse complexities that explore the upper limits of MARL capabilities for real-world applications and reveal key insights.
\end{abstract}

\section{Introduction}
Mass casualty incidents (MCIs) are defined as situations in which casualties greatly outnumber available local resources, which eventually overwhelms the local healthcare system within a time frame \cite{ems_mci}. Some common types of MCIs involve natural disasters such as wildfires, hurricanes, and earthquakes or terrorist attacks such as mass shootings and bombings. When emergency medical responders arrive at the scene of an MCI, an initial task is to locate individuals and assess their injuries promptly, known as triage, followed by physically tagging them with a color-coded label corresponding to injury severity. All victims are tagged before assessing the next steps. Fast and efficient victim tagging is a critical initial stage in the MCI response process, providing vital information that guides subsequent decisions. It enables the timely assessment of the total number of victims, their distribution based on the severity of injuries, and their respective locations throughout the MCI environment.

Most of the current work to improve the emergency response to MCIs focuses on improving resource allocation \cite{Hawe}, responder coordination \cite{Gonzalez, Gonzalez-2}, or transportation to treatment facilities \cite{Wang}. In \cite{Bellamine}, authors model the rescue, treatment, and transportation of MCI victims to hospital locations, testing communication modes between agents. In \cite{tomczyk2023}, authors suggest that timing is a critical factor in saving lives and discuss the triage process; however, similar to other work, they overlook and assume the tagging process. 
There is a lack of research done in the on-scene or pre-hospitalization stage, specifically the task of tagging victims fast. 
To our knowledge, the problem of minimizing victim tagging time has not been formally addressed. Therefore, we leverage insights from MCI response and similar fields, such as search and rescue, to compare and evaluate efficient, practical heuristics for this problem. Our goal is to develop insights that help emergency departments refine MCI planning procedures and enhance the coordination of human or hybrid teams through more effective victim tagging strategies.

Multi-agent reinforcement learning (MARL) has proven effective in a range of domains, such as emergency response and collaborative robotics, because it naturally handles decentralized decision-making under uncertainty \cite{lee2021, zhou2023, yang2020}. Existing MARL research for disaster response, such as fire evacuation or patient admission, typically restricts environments to smaller grids (e.g., 25 $\times$ 25 in \cite{yang2020}) or fewer agents (e.g., up to 100 in \cite{zhou2023}). Nonetheless,  these studies highlight the promise of using reinforcement learning (RL) for complex, evolving tasks like victim tagging.

In this paper, we study the victim tagging problem for a team of medical responders. We formalize victim tagging as an integer linear programming problem and evaluate and compare five distributed, on-the-go responder heuristics to minimize the time it takes to tag all victims under varying levels of uncertainty in communication. 
We further investigate the performance of a MARL method, \textbf{F}actorized \textbf{D}eep \textbf{Q-}\textbf{N}etwork, which we call "FDQN", to minimize victim tagging time, as compared with the heuristic baselines.

Our results show that, among the heuristics, the Local Nearest Victim Policy outperforms others, and local policies (under greater uncertainty) are more efficient than global policies for adaptively choosing the next victim to tag. Individual heuristic policy analyses give us additional insights into the number of responders needed based on the number of victims and the efficiency of each policy, measured in the time it takes to tag victims.
Our evaluation also shows that the proposed FDQN outperforms heuristic baselines in simpler scenarios with lower responder-to-victim ratios and in smaller MCI environments, while heuristics perform better in more complex settings. These results provide valuable insights both for MARL researchers pursuing more realistic, large‐scale applications and for emergency management departments seeking to improve their decision-support strategies.

The contributions of this paper are summarized as follows:
\begin{enumerate}
    \item 
    We present the first integer linear programming (ILP) formulation of the MCI victim tagging task. We expose its combinatorial complexity and use the formulation as a baseline to demonstrate our heuristics and multi-agent reinforcement learning method.
    \item We design five distributed, on-the-go heuristic solutions considering global and local agent communication and analyze their performance through simulations. Results indicate that a local, nearest victim tagging solution minimizes victim tagging time in uncertain situations, as compared to four other practical heuristics.
    \item We design and implement a factorized deep Q-network (FDQN) featuring a shared state space, decentralized decision-making, and constraint-based action pruning for cooperative multi-agent victim tagging. We evaluate its performance against baseline heuristics through extensive, scalable simulation experiments. Results show that FDQN outperformed heuristics in smaller‐scale MCIs and shows promise for scaling to larger environments and higher agent counts. Our experiments have varying scales, some with expansive areas and sizeable agent teams, to explore upper limits of performance, thus offering key insights into real-world MARL applications. 
\end{enumerate}

This article is an extended version of a previous conference paper \cite{CASE_paper}. In this journal article, we propose the use of FDQN 
to learn optimal responder team policies for efficient victim tagging in an MCI emergency medical response. We implemented FDQN and conducted additional experiments, involving our RL strategy integrated with the multi-agent system, as well as additional heuristic-based simulations for comparison. Finally, more technical details and discussions have been added. 

\section{Related Work} \label{RW}
We first survey problems and methods that share parallels with victim tagging and highlight why existing solutions do not fully address the demands of real‐time, uncertain multi‐agent environments. We then explore MARL as an adaptive approach suited to these challenges.

\subsection{Related Problems and Methods}
\subsubsection{Search and Rescue (SAR) as Task Allocation}
Victim tagging has not been widely studied, so we look to SAR literature for similar multi-agent decision-making scenarios. 
Many studies evaluate SAR as a task allocation problem, considering a team of unmanned aerial vehicles (UAVs), hybrid vehicles \cite{Turner, Geng, zhao}, or generally robotic teams \cite{Shi}. 
Typical solutions partition the environment into regions to be explored or tasks to be completed, and then allocate them offline or through a single, centralized planner \cite{Chen, Geng}. For example, in \cite{Geng}, authors use particle swarm optimization, which is still largely centralized in practice. Market‐based auctions such as CBBA (consensus‐based bundle algorithm) \cite{choi2009} or dynamic auctions \cite{Shi} rely on iterative message passing among agents, or on a designated “auctioneer” \cite{Luo}, to coordinate tasks. While these auctions can be partially distributed, they typically assume reliable communication to share bids and converge on consensus.

Other studies choose more distributed approaches, letting agents coordinate with less information. In \cite{zhao}, local "planners" manage subregions. In \cite{Bramblett}, agents physically meet at rendezvous points to exchange new data, which can be infeasible under tight time constraints. Meanwhile, \cite{Pallin} proposes an online genetic algorithm for multi-agent SAR, allowing handovers among agents, but still relies on robust communication of these handover values. A common challenge in these algorithms is their limited adaptivity after the allocation is decided, along with difficulties in fully decentralized communication.
 \subsubsection{Distinctions for Victim Tagging}
Compared to classical SAR, our victim tagging scenario emphasizes \emph{on‐the‐go decision‐making} under uncertain conditions upon arrival. Most prior work focuses on advanced robotic systems or assumes stable global communication. By contrast, we consider both global and local communication and design heuristics that real‐world human responders or hybrid teams can employ. This distributed perspective, combined with more dynamic re‐planning, makes our approach more suitable for high‐uncertainty or large‐scale MCI environments.

\subsection{Multi-agent Reinforcement Learning}
\subsubsection{Deep Reinforcement Learning Motivation}

MCIs are inherently complex and uncertain, requiring adaptive decision-making strategies. RL provides a framework for handling such dynamic environments by enabling agents to iteratively refine their strategies through interaction and feedback. Unlike traditional rule-based approaches, RL can discover optimal solutions without requiring prior knowledge of the environment, making it particularly well-suited for adaptive, high-stakes decision-making scenarios \cite{sutton2018}. Q-learning is a fundamental RL algorithm, which learns an optimal action-selection policy by estimating the expected cumulative rewards (Q-values) for state action pairs \cite{watkins1992}. This method, however, becomes computationally impractical for large state spaces. Deep Q-networks (DQNs) extend Q-learning by using deep neural networks to approximate Q-values, enabling agents to handle high-dimensional state spaces and complex coordination dynamics \cite{mnih2015}. 

\subsubsection{Value Decomposition Networks for Multi-agent Reinforcement Learning}
In a setting where many agents must make decisions, cooperative MARL facilitates coordination among the agents to achieve shared objectives \cite{oroojlooyjadid2019, azadeh2024}. However, MARL faces inherent challenges, including the non-stationarity of simultaneous agent learning, combinatorial explosion of joint actions, and credit assignment, where agents must learn how their individual actions contribute to the overall team reward \cite{oroojlooyjadid2019, azadeh2024, Hernandez-Leal2019,nguyen2020}. Value decomposition is a key strategy to mitigate these issues as it factorizes a global value function into agent‐specific components, enabling each agent to optimize locally while contributing to a shared goal. Value Decomposition Networks (VDNs) introduce a simple additive decomposition of Q-values \cite{Sunehag2017}. QMIX extended this method with a more flexible mixing network to capture agent inter-dependencies \cite{rashid2020}. QTRAN \cite{son2019} and Qatten \cite{yang2020Qatten} further refined the decomposition process, but with high computational costs, complex training processes, and limited real-world validation.

\subsubsection{Our Factorized Approach}
In this work we adopt the basic principle of VDNs. Each responder agent learns an individual Q-function, and these Q-values are then summed to produce a global Q-value for the entire team.  This factorization enables decentralized action selection while providing a shared reward signal to encourage cooperation. Unlike a standard VDN, however, we also assume all responders share a global state to facilitate communication. This allows each agent to see relevant information about the environment and other agents. Additionally, we implement action masking, which prunes the combinatorial action space by disallowing infeasible choices and leveraging real-world constraints for each responder agent. This combination of factorized Q‐learning and domain‐specific action masking allows us to handle the larger environment sizes and higher agent counts characteristic of MCIs while maintaining tractable training complexity.

\section{Problem Description} \label{Prob}
In the victim tagging problem,
the aim is to minimize the total time it takes to tag all victims. Medical responders are referred to as responders, responder agents, or agents throughout the paper, and can be generalized to make up human, hybrid, or fully robotic teams. In this section we set up the preliminaries, and define the problem mathematically.

\subsection{Preliminaries}

 Let us consider a team of $n$ responders $R = \{r_1, \ldots,r_n\}$. A responder $r_i$ is characterized by its speed ($\omega_i$) and the time it takes to tag a victim $v_k$ ($\tau_{ik}$).
The responder agents have a goal to tag all $m$ victims $V = \{v_1,\ldots,v_m\}$. 
We use an $m$ $\times$ $m$ matrix $D$
 to represent the distances between pairs of victims. $D_{jk}$ denotes the distance between victims $v_j$ and $v_k$. If $v_j$ and $v_k$ are the same victim, we assume the distance is 0, i.e. $D_{jk}$ = 0. In this paper we assume all responders start from the same position that we denote with index 0, therefore $D_{0k}$ indicates the distance from the starting point to victim $v_k$.



\subsection{Problem Formulation}
The multi-agent victim tagging problem defined above is a combinatorial optimization problem, which can be formulated and solved using integer linear programming (ILP). The problem is defined as minimizing a linear objective function (the maximum time it takes to tag all victims) subject to linear constraints.

We use $\mathfrak{T}_{ijk}$
to represent the time cost of $r_i$ moving from $v_j$ to $v_k$, represented as 
\begin{equation}\label{TC}
    \mathfrak{T}_{ijk} = \frac{D_{jk}}{\omega_i}.
\end{equation}

The moving paths of each responder can be summarized using a three dimensional matrix $X = \{x_{ijk} | i \in \{1, \ldots,n\}, j \in \{0,\ldots,m\}, k \in \{1,\ldots,m\}\}$. Each element $x_{ijk}$ is a binary variable that represents whether responder agent $r_i$ moves from $v_j$ to $v_k$. The binary values can be defined as

\begin{equation}
    x_{ijk} = \begin{cases} 
1 & \text{if } r_i \text{ moves from } v_j \text{ to } v_k, \\
0 & \text{otherwise} .
\end{cases}
\end{equation}
Note that $j$ and $k$ represent the victim index, except in the case where $j = 0$, which indicates that $r_i$ is at the starting location ($k \neq 0$ since responder agents do not need to return to their starting location).

Then the victim tagging problem can be formulated as the following minmax ILP:
\begin{equation} \label{Obj}
    Obj = \min \left\{ \max_{1 \leq i \leq n} \left\{ \sum_{j=0}^{m} \sum_{k=1}^{m} (\mathfrak{T}_{ijk} + \tau_{ik}) x_{ijk} \right\} \right\},
\end{equation}

where $Obj$ is the objective function describing the goal to minimize the maximum time it takes for all responders to tag victims. It is subject to the following constraints:

\begin{equation} \label{five}
    \sum_{i = 1}^{n} \sum_{k = 1}^{m} x_{i0k} \leq n
\end{equation}

Equation (\ref{five}) specifies that at most $n$ responders leave the starting position. 

\begin{equation}
\label{six}
\sum_{i = 1}^{n} \sum_{j = 0}^{m} x_{ijk} = 1, \quad \forall k \in \{1, \ldots, m\}
\end{equation}

Equation (\ref{six}) shows that exactly one responder tags each victim $v_k$.

\begin{equation} 
\label{seven}
    \sum_{i = 1}^{n} \sum_{k = 1}^{m} x_{ijk} \leq 1, \quad \forall j \in \{1, \ldots,m\}
\end{equation}

Equation (\ref{seven}) specifies that at most one responder leaves each victim $v_j$. When all victims are tagged, responders remain at the last victim they tag.

\begin{equation} 
\label{eight}
    \sum_{j = 0}^{m} \sum_{k = 1}^{m} x_{ijk} \leq m, \quad \forall i \in \{1, \ldots,n\}
\end{equation}

Equation (\ref{eight}) illustrates that each responder $r_i$ tags at most $m$ victims.

\begin{equation} \small
\label{nine}
\sum_{j=0}^{m} x_{ijk} + \sum_{\substack{a=1 \\ a \neq i}}^{n} \sum_{b=1}^{m} x_{akb} \leq 1,\forall i \in \{1, \ldots, n\}, \forall k \in \{1, \ldots, m\}
\end{equation}
Equation (\ref{nine}) ensures that if a responder $r_i$ tags victim $v_k$, then only $r_i$ can leave $v_k$. 

\begin{equation} \label{thirteen} 
    x_{ijk} \in \{0,1\}, \quad \forall i \in \{1, \ldots,n\}, \forall j \in \{0, \ldots,m\},
\end{equation}
\begin{equation*} 
    \forall k \in \{1,\ldots,m\}
\end{equation*}

Equation (\ref{thirteen}) describes that the path of $r_i$ from $v_j$ to $v_k$ is either 0 or 1.

\begin{equation}
\label{ten}
    u_{i0} = 0 \quad \forall i \in \{1,\ldots,n\}
\end{equation}

\begin{equation}
\label{eleven}
    1 \leq u_{ik} \leq m \quad \forall i \in \{1,\ldots,n\}, \forall k \in \{1,\ldots,m\}
\end{equation}

\begin{equation}
\label{twelve} \small
    u_{ij} - u_{ik} + 1 \leq 
    m(1-x_{ijk}), \quad \forall i \in \{1,\ldots,n\}, \forall j \in \{0,\ldots,m\},
\end{equation}
\begin{equation*} \small
    \forall k \in \{1,\ldots,m\}
\end{equation*}

\begin{equation} \label{fourteen} 
    u_{ij} \in \{0,\ldots,m\}, \quad \forall i \in \{1, \ldots,n\}, \forall j \in \{0, \ldots,m\}
\end{equation}

Equations (\ref{ten} - \ref{fourteen}) ensure that the paths of the $n$ responders do not contain cycles (i.e. they do not return to a previously tagged victim). Here, we extend the Miller-Tucker-Zemlin (MTZ) formulation for the Traveling Salesman Problem \cite{MTZ,mtz2} to $n$ paths. Equation (\ref{fourteen}) specifies that $u_{ij}$ from the MTZ-based formulation accounts for all $n$ responders and $m$ victims.



\section{Approach} \label{approach}
Equations (\ref{Obj} - \ref{fourteen}) describe the victim tagging problem assuming optimal conditions, with centralized, global information available and optimal communication within the responder team. 
It considers all possible victim tagging cases for each responder, which is laborious and time-consuming. Solving this type of problem optimally is computationally intractable due to its combinatorial nature and NP-Hard complexity
\cite{Geng}. 
The complexity of the problem grows exponentially as the number of agents increases, thus an optimal solution is not suitable for real-time practical applications. Additionally, in a real MCI scenario, global information and optimal communication are not always available. Therefore, we first formalize and present baseline heuristic solutions that account for varying degrees of uncertainty. 
Then, since adaptive solutions are advantageous for this problem, we formalize and present a factorized deep Q-network (FDQN). We evaluate the effectiveness of FDQN in generating optimized policies to accelerate victim tagging compared to heuristic approaches.

\subsection{Baseline Heuristics}
For the baseline heuristics, we focus on the practical aspect of MCI victim tagging and consider solutions (referred to as policies) that are commonly used in emergency response or related fields. Unlike other heuristics such as CBBA \cite{Turner, Luo}, 
particle swarm \cite{Geng}, and genetic algorithms \cite{Pallin}, ours are practical and intuitive approaches that are used in practice for MCIs or could easily be applied by responder teams in chaotic, stochastic MCI scenarios on the go without having extensive prior knowledge of the situation. The five heuristics we explore are based on existing methods, but, to our knowledge, they have not been formally defined and quantitatively compared to optimize victim tagging time.


Each policy we present acts as the overarching responder team policy, which allows for responder agents to individually and adaptively select their next victim to tag. For the following policy descriptions, we refer to the previously defined term $x_{ijk}$, which describes whether responder agent $r_i$ moves from a victim with index $j$ to a victim with index $k$.
Initially all $x_{ijk} = 0$ and $j = 0$ for all responders. The responder team's policy will determine how an individual responder locally and iteratively determines $k$, which then becomes $j$ when moving to the next victim to tag. The selection of $k$ for each responder agent happens on the go, individually for each responder, in a distributed manner. Therefore, when victim tagging is completed, a sequence of $k$ values determines the order of tagged victims' indices for one responder, and $k = j + 1$ in the sequence of victims tagged. 

To explain our heuristics, we introduce responder attribute $u_i$, which is initialized as Null, then becomes the victim that $r_i$ has selected to tag next, and continues to update as $r_i$ tags victims based on the policy. A victim $v_j$ is characterized by the responder that is going to tag them ($f_j$) where $f$ is the responder ID, and whether they have been tagged yet ($g_j$), where $g_j \in \{\{0,1\} | 0 = \text{untagged}, 1 = \text{tagged}\}$. This section formally presents the five policies we analyze as methods for a responder $r_i$ to identify the next victim to tag ($v_k$). For any definition of $k$, if $k$ is undefined, then $k$ = $j$ indicating the responder is idle. For all formalizations, dist($a, b$) refers to the Euclidean distance between agents $a$ and $b$.

\subsubsection{Random Victim Policy (RVP)}
In RVP, a responder $r_i$ chooses a random victim $v_l$ that has not been tagged yet, and one who is not chosen by a different responder to tag. Using previously defined responder and victim attributes, the victim's index can be mathematically formulated as

\begin{equation}
\begin{aligned}
    k &= \{\text{random } v_l \in V \, | \, ((v_l \notin u_i , \forall r \in R) \wedge \, g_l = 0)\}.
\end{aligned}
\end{equation}
RVP represents the tagging method that assumes all globalized and centralized information with minimal uncertainty and, therefore, acts as our benchmark. The four following policies build on this one and allow for increasing uncertainty. 


\subsubsection{Nearest Victim Policy (NVP)}
In NVP, a responder $r_i$ chooses the nearest victim that has not been tagged yet and is not chosen by a different responder as its target victim to tag. This is represented as

\begin{equation}
\begin{aligned}
    k &= \text{arg min}_{v_l \in V}\{\text{dist}(r_i, v_l) \, | \, ((v_l \notin u_i, \forall r \in R) \wedge \, g_l = 0)\}.
\end{aligned}
\end{equation}
NVP is loosely based on a practical heuristic in \cite{Bellamine}. This policy continues to be applicable in situations that allow for global communication and perception.

\subsubsection{Local Nearest Victim Policy (LNVP)}
LNVP describes the situation where responder $r_i$ locally chooses the nearest victim as its next victim to tag. Responder $r_i$ can choose a victim that is previously chosen by a different responder, in which case the prior responder may need to replan its next steps based on this on-the-go iterative process. 
This is formulated below:
\begin{equation} 
\begin{aligned}
    k &= \text{arg min}_{v_l \in V}\{\text{dist}(r_i, v_l) \, | (g_l = 0 \wedge (f_l = \text{NULL} \\ &\vee (\text{dist}(v_l,f_l) > \text{dist}(r_i,v_l) \wedge \text{dist}(v_l,f_l) > \zeta)))\},
\end{aligned}
\end{equation}
where $\zeta$ acts as a threshold and can be any value or function indicating how far a victim's current tagger has to be from the victim so that they can reroute if needed. LNVP is similar to NVP, except it is a heuristic that allows for increased uncertainty. In LNVP, we assume responders do not have the ability to communicate globally, only locally. 

\begin{table}[t]
  \centering
  \footnotesize
  \caption{Important variables used to formalize our heuristics (left) and FDQN (right).}
  \begin{tabular}{p{0.65cm}p{3.2cm}|p{0.65cm}p{3.2cm}}
    \toprule
    Symbol & Description & Symbol & Description \\
    \midrule
    $R$ & Set of responder agents  & $\mathcal{S}$ & Global state space \\
    $n$ & Number of responders  & $\mathcal{A}$ & Combined action space \\
    $r_i$ & The $i$th responder & $P$ & Transition function \\
    $u_i$ & Selected victim for $r_i$  & $\mathcal{R}$ & Reward function \\
    $\tau_{ik}$ & Time it takes $r_i$ to tag $v_k$ & $\gamma$ & Discount factor \\
    $\omega_i$ & Speed of $r_i$ & $B$ & Number of bins \\
    $c_i$ & Cell for $r_i$ & $d_{ij}$ & Binned distance ($r_i$ to $v_j$) \\
    $\psi_i$ & State of $r_i$ in FSM & $\xi_j$ & Selection status for $v_j$ \\
    $V$ & Set of victim agents &  $\pi$ & Sequence of actions\\
    $m$ & Number of victims & $Q$ & Action-value function \\
    $v_j$ & The $j$th victim & $\mathcal{D}$ & Replay buffer \\
    $h_j$ & Health status of $v_j$ & $y$ & Target network \\
    $f_j$ & The responder that selected $v_j$&\(f_{\mathrm{update}}\) &  Target network update frequency \\
    $g_j$ & Tagged status for $v_j$ 
     &  $\theta$ & Network parameters \\ 
    $p_j$ & Position of $v_j$ &  $\theta'$ & Target network parameters \\
    $\zeta$ & Threshold for distance value &   &  \\
    \bottomrule
  \end{tabular}
  \label{tab:notation}
\end{table}


\subsubsection{Local Critical Victim Policy (LCVP)}
In LCVP we introduce a new victim attribute $h_j$, which indicates victim $v_j$'s health status. This can be defined using any criteria, but we represent it following the START (Simple Triage and Rapid Treatment) algorithm, which is commonly used in MCIs to correspond triage tags to health status using colors \cite{START}. For $v_j$, if \(h_j \in [0, 1]\), we can denote (1) Black (dead) when \(h_j \in [0, 0.25)\), (2)  Red (immediate) when \(h_j \in [0.25, 0.50)\), (3) Yellow (delayed) when \(h_j \in [0.50, 0.75)\), and (4) Green (mobile) when \(h_j \in [0.75, 1]\). Health status is an internal measure to the victim, so we assume that if a responder sees a victim, it can visually determine whether the victim is critically injured (dead or almost dead) or less injured (mobile or slightly injured). Thus, for LCVP we define a critical victim as a victim with attribute $h_j < 0.5$.

For LCVP, a responder $r_i$ chooses to tag the nearest critical victim and then follows LNVP when all critical victims are tagged. This is similar to \cite{Bellamine}, where they utilize memory-guided finding,
except that in our case, the responder agents are not aware of the exact injury severities and have not previously seen the MCI environment. This is formally defined here:
\begin{equation}
\begin{aligned}
    k &= \text{arg min}_{v_l \in V}\{\text{dist}(r_i, v_l) \, | \, (g_l = 0  \wedge h_l < 0.5 \\ & \wedge  (f_l = \text{NULL} \vee \, (\text{dist}(v_l,f_l) > \text{dist}(r_i,v_l) \\ &\wedge \, \text{dist}(v_l,f_l) > \zeta)))\},
\end{aligned}
\end{equation}

\begin{equation} \label{LCVPpt2}
\begin{aligned}
&\text{if } (\neg \exists v_l | h_l < 0.5) \wedge (\exists v_l | g_l = 0) \text{, then use } LNVP.
\end{aligned}
\end{equation}

Here, $\zeta$ acts the same as in LNVP.
Equation (\ref{LCVPpt2}) indicates that after all critical victims are tagged, and there are still victims that have not been tagged, then $r_i$ will follow LNVP to find $k$, the index of the next victim they should tag. For LCVP we assume responders do not have the ability to communicate globally, they must communicate ad hoc locally.

\subsubsection{Local Grid Assignment Policy (LGAP)}
For LGAP, the MCI environment is divided into cells, so that each responder is assigned to a cell in the grid. For this policy, we introduce a set of cells $C = \{c_1, \ldots,c_n\}$ describing cell areas for $n$ responders, a responder attribute $c_i$, and a victim attribute $p_j$. In LGAP, a responder $r_i$ tags the nearest victim within its cell $c_i$ until all victims in its cell are tagged. A victim $v_j$'s position in the MCI environment is represented as $p_j$. LGAP is formalized as follows:

\begin{equation}
\begin{aligned}
    k &= \text{arg min}_{v_l \in V}\{\text{dist}(r_i, v_l) \, | (p_l \in c_i \wedge g_l = 0 )\}.
\end{aligned}
\end{equation}

We designed LGAP based on commonly used practical SAR heuristics where different specified areas should be explored \cite{Chen}. For each aforementioned policy we assumed that responders have a global view of the MCI environment and the victims within it. LGAP allows for the most uncertain conditions, where only local communication and limited perception exist. We assume that the perception of a responder agent is equal to or greater than its cell size. The variables used to summarize our heuristics are summarized in Table \ref{tab:notation}.

\subsection{Factorized Deep Q-Network (FDQN)}

Given the complex and uncertain environment of an MCI, we adopt a MARL approach to optimize policy generation for victim tagging. In particular, we implement FDQN, where each responder agent acts individually, selecting from its own action space, then a global Q-value is computed by summing each agent’s contribution. This factorization sidesteps the exponential growth of a fully joint action space and provides shared feedback for cooperative behavior. This assumes that the responder agents have global communication, represented by a shared state space.


\begin{figure}[t]
    \centering
    \includegraphics[width=0.4\textwidth]{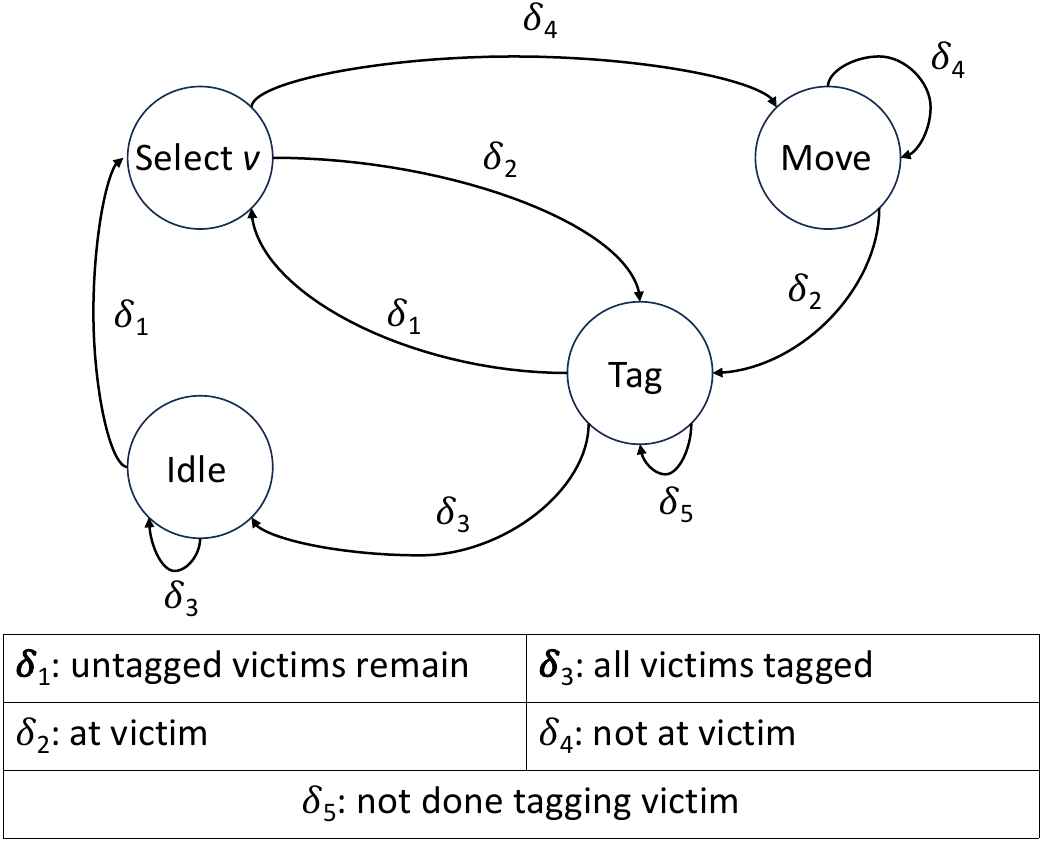}
    \caption{Responder agent finite state machine (FSM). The state 'select $v$' indicates selecting a victim.}
    \label{fig:fsm}
\end{figure}

\subsubsection{MARL Setting}
The victim tagging problem can be cast as a multi-agent Markov Decision Process (MDP) defined by the tuple \( (n, \mathcal{S}, \mathcal{A}, P, \mathcal{R}, \gamma)\), where $n$ is the number of responder agents, \( \mathcal{S} \) is the global state space, \( \mathcal{A} \) is the combined action space of all agents, \( P(s'\mid s,a) \) is the stochastic transition function, \( \mathcal{R}(s, a) \) is the reward function, and \( \gamma \in (0,1] \) is the discount factor.
These are defined as follows:
\begin{itemize}
    \item \textbf{Agent:} Each responder $r_i$ is considered a learning agent, with the set of responders $R = \{r_1, \ldots,r_n\}$ representing all agents.
    \item \textbf{State}: A global state $s_t$ at time $t$ is defined as a finite set: 
\begin{equation}
\small
   s_t = [d_{1,1},\dots,d_{n,m},\;
   \psi_1,\dots,\psi_n,\;
   \xi_1,\dots,\xi_m,\;
   g_1,\dots,g_m].
\end{equation}
    This consists of binned distances between each pair of responders and victims $d$, responder states $\psi$, victim selection statuses $\xi$, and victim tagged statuses $g$.    
    These are defined as follows:
    \begin{itemize}
    \item \textit{Binned Distances}: 
    To facilitate informed decision-making among agents, we track the locations of agents using pairwise distances. The distances between each pair of responders and victims are discretized into $B$ bins to reduce the state space while maintaining approximate spatial relationships. 
    The binned distance $d_{ij}$ between a responder \( r_i \) and victim \( v_j \) is defined as the following:
\[
\footnotesize
d_{ij} =
\begin{cases}
   0, & \quad \text{if } \quad \text{dist}(r_i,v_j) < \zeta,\\[5pt]
   1, & \quad \text{if } \quad 
   \begin{aligned} 
       \zeta &\leq \text{dist}(r_i,v_j) \\ 
       &< 2\zeta
   \end{aligned},\\[5pt]
   \min\!\Bigl( \bigl\lfloor \frac{\text{dist}(r_i,v_j)}{w/B} \bigr\rfloor, B-1 \Bigr), & \quad \text{otherwise.}
\end{cases}
\]
Variable \( w \) is the width of the MCI area, \( B \geq 2 \) is the number of bins,  and $\zeta$ is a threshold which can be any value or function to indicate closeness. We track a binned distance $d_{ij}$ for each pair of responders and victims.
\item \textit{Responder States}: Each responder agent follows a finite state machine (FSM) that governs its internal behavior and progression through tasks (e.g., moving to a victim before tagging). For this victim tagging problem, we design a FSM shown in Fig. \ref{fig:fsm}. The FSM structure constrains which actions are feasible at any time step, ensuring consistent decision‐making. Formally, let 
$
\psi = \{\psi_i \mid r_i \in R\}, 
\quad 
\psi_i \in \{\text{idle}, \text{select $v$}, \text{move}, \text{tag}\},
$
where \( \psi_i \) denotes the current FSM state of responder \( r_i \) and $\lvert\psi\rvert$ is the total number of FSM states. We track each responder’s state \(\psi_i\) as part of the global state.

\item \textit{Victim Selection Statuses}: When a responder chooses a victim for tagging, that victim is marked as “selected.” We encode this as \(\xi_j \in \{0,1\}\) for each victim \(v_j\), where \(\xi_j = 1\) if \(v_j\) is currently selected by some responder, and \(\xi_j = 0\) otherwise.  We track the selection status $\xi_j$ for each victim.

\item \textit{Victim Tagged Statuses}: The representation of victim tagged statuses is as follows: \( g_j \in \{0,1\} \), where \( g_j = 1 \) if victim \( v_j \) has been tagged, otherwise \( g_j = 0 \). We track the tagged status $g_j$ for each victim.
\end{itemize}
\item \textbf{Action}:
Each agent $r_i$ selects an individual action from its action space $\mathcal{A}_i$ at time $t$. 
The action space for each agent includes one action corresponding to each FSM state, except for the 'select $v$' state. Instead, 'select $v$' is expanded into $m$ victim-selection actions (one for each victim). Thus, if $\lvert\psi\rvert$ is the number of FSM states and $m$ is the number of victims, then each agent's action space is as follows:
\begin{equation}
\mathcal{A}_i = \{ a_0, \dots, a_{\lvert\psi\rvert+m- 2}\},
\end{equation}
where the first $\lvert\psi\rvert - 1$ actions correspond to the remaining FSM states, and the final $m$ actions correspond to selecting each victim.
Each individual action is further described in Section \ref{sec:sim_setup}.
A shared policy $\pi^\ast$ emerges from all agents picking actions, or sub-actions, simultaneously, forming a joint action $
a_t = \{ a_{1(t)},\dots, a_{n(t)} \}, \text{where}\quad a_{i(t)} \in \mathcal{A}_i, \quad \forall r_i \in R$. 

We enforce the responder FSM by applying\textit{ action masking}, setting
$Q_i(s, a_i) = -\infty$
for any choice that conflicts with the current FSM state. For instance, if an agent is mid-movement and has not reached its victim, the only valid action is to continue moving. Similarly, if all victims are tagged except one currently being handled, other responders must idle rather than re-select that victim. This ensures each agent strictly follows the FSM transitions for victim tagging and substantially reduces the effective action space, thereby accelerating FDQN convergence.
\\
To enhance knowledge sharing and prevent redundant efforts, we implement a retrospective conflict-resolution procedure. After all agents select their actions at each time step, this procedure resolves conflicts before execution. For example, if multiple agents select the same victim, the agent less likely to benefit (e.g., the one farther away) defaults to an 'idle' action.

\item \textbf{Discount Factor} $\gamma$: The discount factor $\gamma \in (0,1]$ quantifies the difference in prioritization between immediate and future rewards for decision making. 


\item \textbf{Reward Function}: 
Each responder $r_i$ obtains an individual reward $\mathcal{R}_{i(t)}$ at each time step, and these rewards sum to $\mathcal{R}_{\text{total}(t)} = \sum_{r_i \in R} \mathcal{R}_{i(t)}$. We encourage efficient victim tagging and penalize slow progress or unnecessary actions. For each individual agent, a penalty of -1 is obtained at each time step. However, when a responder successfully tags a victim, they receive a reward
    \begin{equation}
        \mathcal{R}_{i(t)} = \mathcal{R}^b \times \left(1 + 0.1 \times V_{tagged(t)}\right),
    \end{equation}
    where \( V_{tagged(t)} \) represents the total number of victims tagged at time \( t \). Here, the base reward \( \mathcal{R}^b \) is empirically derived and defined as:
    \begin{equation}
        \mathcal{R}^b = 30 - 0.5 \times \left\lfloor \frac{\text{steps}_t}{10} \right\rfloor,
    \end{equation}
    which introduces a progressive penalty for every 10 time steps, encouraging responders to tag victims efficiently.
    




\end{itemize}

\subsubsection{Learning Process}
Our objective is to find an optimal joint policy $\pi^*$ that maximizes the cumulative rewards. Each agent $r_i$ learns an action-value function $Q_i$ via Q-learning updates. In the simplest form,
\begin{equation}
Q_i(s,a_i)
\; \leftarrow \;
Q_i(s,a_i)
\;+\;
\alpha \Bigl[
   \mathcal{R}_i
 + \gamma \,\max_{a_i'}\,(Q_i(s',a_i'))
 - Q_i(s,a_i)
\Bigr],
\end{equation}

where $\alpha$ is the learning rate, $\mathcal{R}_i$ is agent $r_i$'s reward, and $s'$ is the next state. Following the VDN-style factorization \cite{Sunehag2017}, we sum individual Q-values to form a joint $Q$:
\begin{equation} \label{eq:Qjoint}
    Q_{\text{joint}}(s,a)
=\;
\sum_{i=1}^n
Q_i(s,a_i),
\end{equation}

with $Q \rightarrow Q^\ast$ as training converges. The factorization is beneficial for the cooperative task of victim tagging, allowing agents to share a global reward while individually selecting optimal actions. 

\subsubsection{Deep Q-Network Approximation}
To handle high-dimensional states and actions, we approximate each agent's action-value function $Q_i(s,a_i)$ with a deep neural network. 

\paragraph{Training Stabilization}
Following the standard DQN paradigm \cite{mnih2015}, we use two primary mechanisms to stabilize training, a replay buffer and a target network.
We maintain a global \textit{replay buffer} $\mathcal{D}$ that stores transitions ($s$, ($a_1, \dots, a_n$), $\mathcal{R}$, $s'$) from all agents. This buffer improves sample efficiency and reduces correlation between consecutive samples by uniformly sampling mini-batches (size 64). We fix its capacity (size 10,000) in a first-in-first-out manner.
We keep a separate \textit{target network} (parameters $\theta'$) that is updated less frequently than the main network (parameters $\theta$). By periodically copying model weights $\theta \rightarrow \theta'$, we prevent the network from chasing a moving target and reduce instability in Q-value estimates.


\paragraph{Temporal Difference Loss}

We train the DQN by minimizing the temporal-difference (TD) loss at each training step:
\begin{equation}
\mathcal{L}(\theta)
\,=\,
\mathbb{E}_{(s,a,\mathcal{R},s') \sim \mathcal{D}}\bigl[
   \bigl(y - Q_{\text{joint}}(s,a;\theta)\bigr)^2
\bigr],
\end{equation}
\begin{equation}
\text{where }
y
\,=\,
\mathcal{R}
\;+\;
\gamma \,\max_{a'}\,
Q_{\text{joint}}\!\bigl(s', a'; \theta'\bigr).
\end{equation}
Here, $\theta$ are the main network’s parameters and $\theta'$ are the fixed target network's parameters. After computing the loss, we perform gradient step via stochastic gradient descent (SGD) with the Adam optimizer.

\paragraph{Factorized Q-Network Architecture}

\begin{figure}[t]
    \centering
    \includegraphics[width=0.48\textwidth]{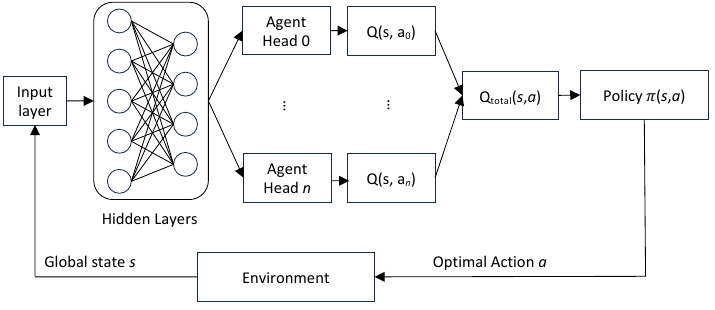}
    \caption{Factorized deep Q-network (FDQN) architecture.}
    \label{fig:DQN_arch}
\end{figure}

We implement a factorized DQN (Fig. \ref{fig:DQN_arch}) by first encoding the global state $s$ through a shared series of fully connected layers. We use two layers with sizes 128 and then 64, each with ReLU activation. Then we split into $n$ agent-specific heads, each outputting $Q_i(s,a_i)$. Summing these heads forms the global Q-value as in Equation (\ref{eq:Qjoint}).
This design preserves decentralized action selection, while still leveraging a shared global state.

\paragraph{Training Routine}
We initialize an environment with $n$ responders and $m$ victims, then run full episodes until all victims are tagged. Each agent picks actions via an $\epsilon$-greedy policy to balance exploration and exploitation. After each step we store ($s$, ($a_1, \dots, a_n$), $\mathcal{R}$, $s'$) in the replay buffer. Every $\mathcal{K}$ steps, we sample a mini-batch and update parameters via the TD loss. Periodically, we sync the target network with the main network. Repeating this across many episodes refines the Q-values until the FDQN converges to a near-optimal policy that efficiently tags victims.

\section{Experiments}\label{exper}

\subsection{Simulation Setup} \label{sec:sim_setup}
For our experiments, we designed a multi-agent system (MAS) consisting of responder agents and victims in an MCI environment, and created simulations utilizing agent-based modeling to extensively test the responder heuristics and FDQN approach. We programmed the simulations in Python (v. 3.11.6) and used the open-source agent framework Mesa (v. 2.2.4) \cite{mesa} for the MAS design. 
The MAS environment has continuous space, runs in discrete time steps, and utilizes a scheduler that activates agents in a random order at each time step. The simulation iterates forward by one time step until all victims are tagged.

The MCI environment is staged as a 2-dimensional rectangular space with area $A$.
The responder agents tag victims following the heuristic policy or learned FDQN policy until all victims are tagged. When victims are tagged by a responder, they are marked as black (dead), red (immediate), yellow (delayed), or green (mobile) following the START response algorithm \cite{START}. For simplicity, specific tags representing victim health are not utilized in the FDQN implementation, but could be for future work. Victim positions are randomly and uniformly distributed throughout the MCI environment, and we assume they remain stationary.


\begin{table}[t] \footnotesize
  \centering
  \caption{Simulation parameters, where $t$ is time and values are arbitrarily chosen.}
  \begin{tabular}{p{3cm}p{1cm}p{2.6cm}} 
    \toprule
    Parameter & Symbol & Value \\
    \midrule
    MCI Area & $A$ & 100 x 60 units \\
    Time to tag any victim & $\tau$ & 3 time steps \\
    Victim health state & $h$ & random value \(\in [0, 1]\) \\
    Victim position & $p$ & random value $\in$ $A$ \\
    Responder speed & $\omega$ & 1 unit/time \\
    Responder start position & $p_{t=0}$ & $0$ \\
    Responder start FSM state & $\psi_{t=0}$ & idle \\
    \bottomrule
  \end{tabular}
  \label{tab:sim_parameters}
\end{table}

For the purpose of our experiments, we use our previously defined FSM responder states for each responder $r_i$, $\psi_i \in$ \{${\text{idle, select $v$, move, tag}}$\} (Fig. \ref{fig:fsm}). 
All responder agents begin in the 'idle' state and then transition between the 'select $v$', 'move', and 'tag' states, following transitions denoted with $\delta$. 
When transitioning to the 'tag' state, a responder $r_i$ is in the process of tagging a victim $v_k$ and is done after $\tau_{ik}$ time steps.
When transitioning to the 'select $v$' state, the agent selects a victim based on its heuristic or learned FDQN policy. When transitioning to the 'move' state, this represents a discrete action where the responder moves towards its selected victim, based on its speed $\omega$.
The attribute $p_{i(t)}$ for an agent additionally describes its position at any given time $t$. Table \ref{tab:sim_parameters} summarizes our simulation parameters for evaluating the baseline heuristics, which can be easily altered based on simulation goals. The FDQN simulation parameters are identical, except for the MCI environment area, which are detailed in Section \ref{sec:FDQN_eval}.

\paragraph{FDQN-specific Setup}
We manually implemented the FDQN and utilized one NVIDIA A6000 GPU for training.
Unlike the heuristic-based approaches, the FDQN does not enforce a predefined responder policy when responders select a victim for tagging. Instead, the responders autonomously learn an effective strategy over the course of training, refining their decision-making through thousands of iterations across diverse scenarios. After training concludes, we evaluate the resulting team policy to assess its performance under the learned decision‐making framework.


\subsection{Between Baseline Heuristics Evaluation}


\begin{table}
  \centering
  \small
  \setlength{\tabcolsep}{2.8pt} 
  \caption{Number of responder (No. R) agents and victims (No. V) for each experiment to evaluate the baseline heuristics. Results highlight the average time steps to tag all victims for each experiment, over fifty iterations. Bold values highlight the most efficient policy for each experiment.}
\begin{tabular}[t]{cc*{8}{c}} 
    \toprule
    \multicolumn{1}{c}{\thead{\small Experiment}} & 1 & 2 & 3 & 4 & 5 & 6 & 7 & 8 & 9 \\
        \addlinespace[-2pt] 
        No. R & 5 & 5 & 5 & 5 & 20 & 20 & 80 & 80 & 320 \\
    No. V & 10 & 20 & 100 & 1000 & 100 & 1000 & 100 & 1000 & 1000 \\
        \midrule

      &  &   \multicolumn{5}{c}{\thead{\small Results (average time steps)}} & &  &  \\

    
    RVP & 136 & 226 & 956 & 9,135 & 299 & 2,381 & 122 & 657 & 231\\
    NVP & 124 & 152 & 298 & 1,328 & 164 & \textbf{411} & 132 & 218 & 152\\
        \cmidrule(lr){1-10}

     LNVP & 118 & \textbf{145} & \textbf{288}  & \textbf{1,316} & \textbf{153} & 420 & \textbf{111} & \textbf{190} & \textbf{118} \\
    LCVP & \textbf{115} & 174 & 375 & 1,573 & 197 & 506 & 117 & 258 & 136\\
    LGAP & 126 & 161 & 317 & 1,383 & 177 & 470 & 148 & 238 & 146 \\
    \bottomrule
\end{tabular}
  \label{tab:results_table}
\end{table}

\begin{figure}[t]
    \centering
    \includegraphics[width=0.486\textwidth]{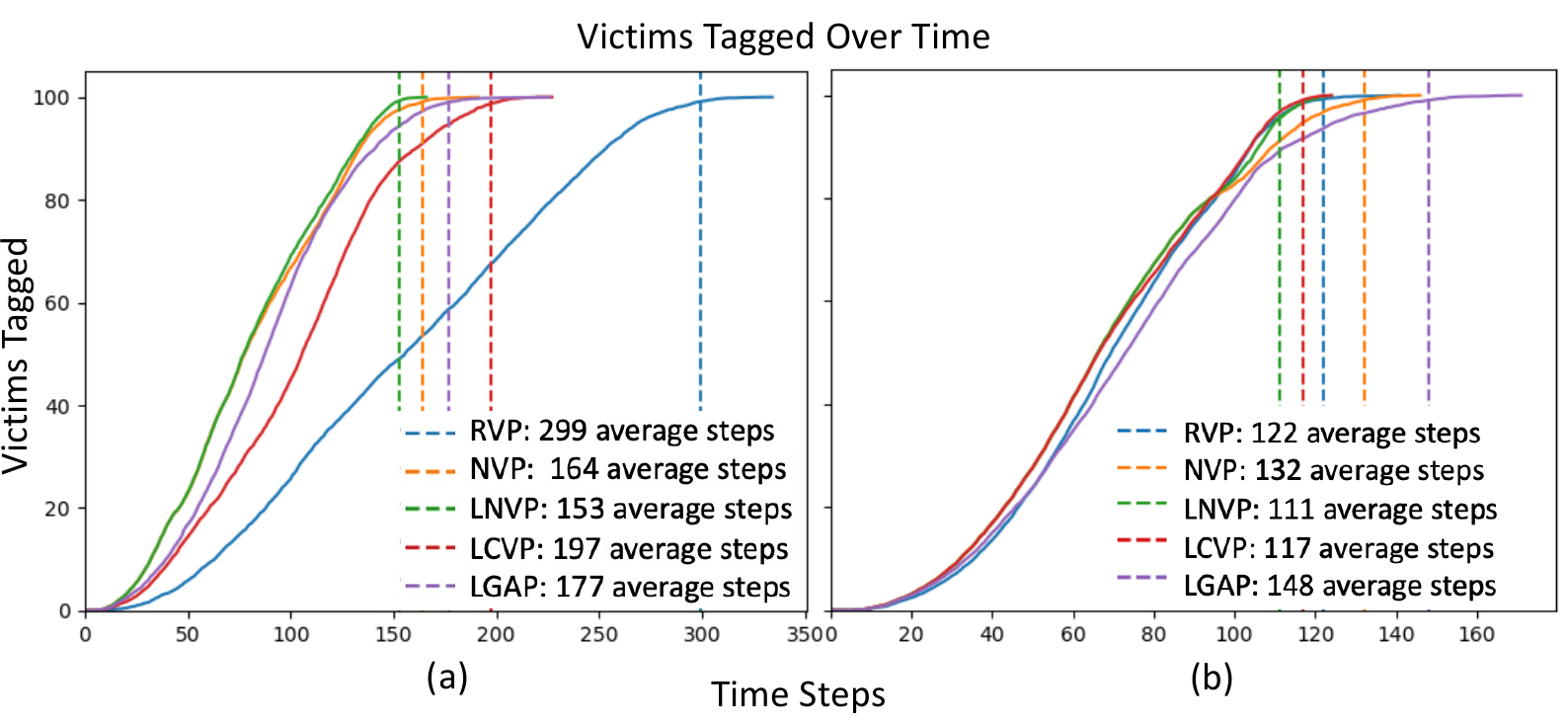}
    \caption{Heuristic policy comparison for the number of victims tagged over each time step for 100 victims and (a) 20 or (b) 80 responders. Each color curve denotes a different policy, and the dotted vertical line shows the average time it takes to tag all victims for each policy.}
    \label{fig:2graphs}
\end{figure}

\begin{figure*}[ht]
    \centering
    \includegraphics[width=\textwidth]{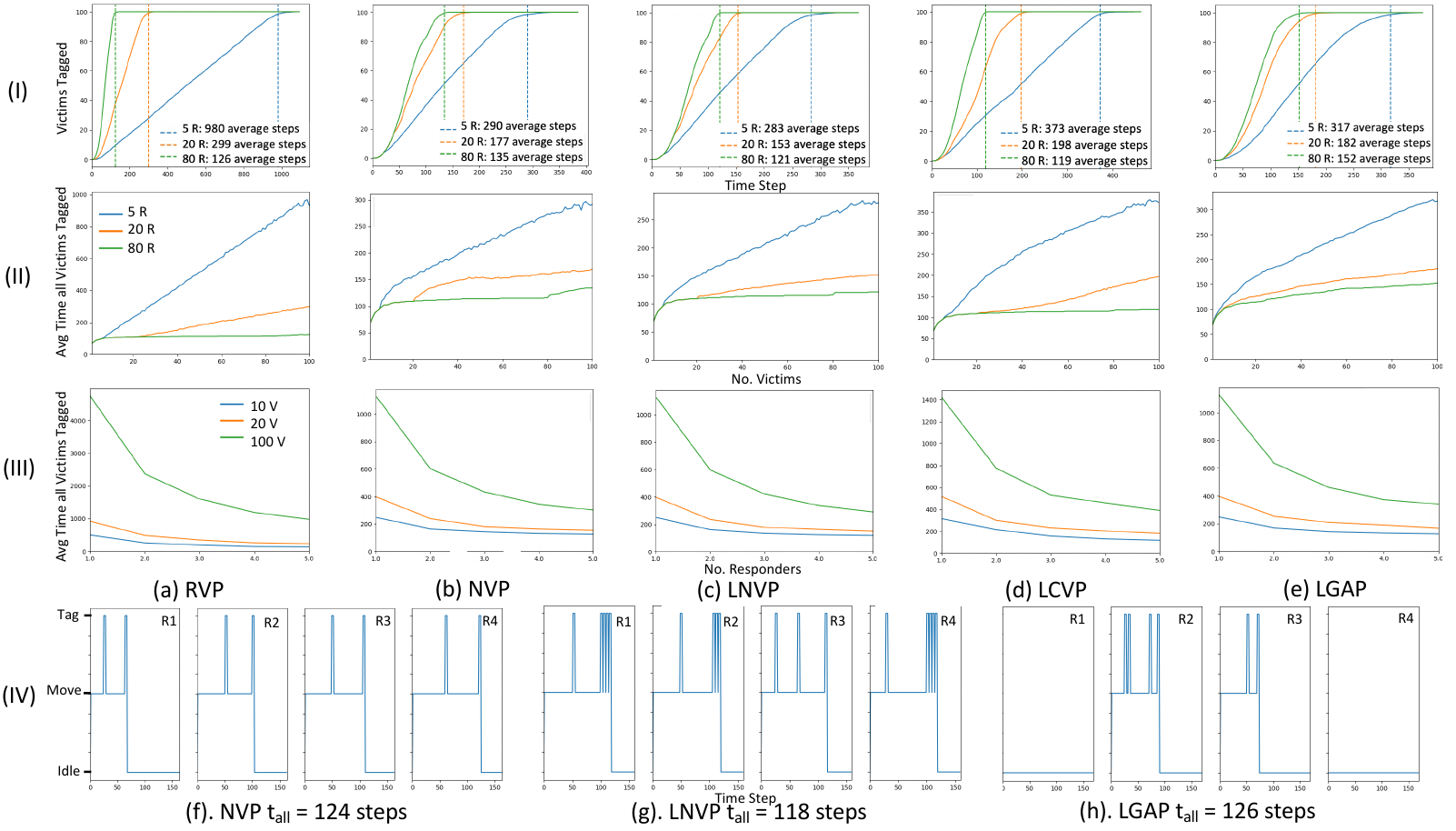}
    \caption{Graphs show results evaluating heuristics. (I) Number of victims tagged over time for 5, 20, and 80 responders. (II) Average time all victims are tagged for up to 100 victims for 5, 20, and 80 responders. (III) Average time all victims are tagged for 10, 20, and 100 victims for up to 5 responders. Columns (a-e) illustrate different policies. (IV) Responder agents' states over time. Four responders are shown for experiment 1 (5R, 10V). Parts (f-h) demonstrate different policies and the time when all victims are tagged ($t_{all}$).}
    \label{fig:all}
\end{figure*}

We devise experiments with varying numbers of agents to represent nine MCI scenarios of different scales to evaluate which heuristic strategy optimally minimizes victim tagging time, shown in Table \ref{tab:results_table}. The number of victims selected is based on one MCI incident plan \cite{REMSC}, and we choose the number of responders to be fewer than the number of victims. For each of these nine experiments, we test each of the five policies for a total of forty-five experimental scenarios. 
For each experiment, we run fifty iterations and report the average results. Each experiment is evaluated using the objective function in Equation (\ref{Obj}) to find the minimal time it takes to tag all victims.

Table \ref{tab:results_table} reports the average number of time steps it took to tag all victims for each experiment, comparing heuristics. Policies with global communication between responder agents (RVP and NVP) are distinguished from policies with local communication (LNVP, LCVP, and LGAP). LNVP performed optimally, with the minimum average time step values across almost all experiments, while RVP performed the least efficiently. 

Each policy's average time to tag all victims increases as the ratio between the number of responder and victim agents grows. We notice RVP's values elevate increasingly compared to other policies. 
In the tests involving five responders (experiments 1-4), as the number of victims increases, the RVP values rise from 136, 226, 956, to 9,135, respectively, while the values for the other scenarios remain lower. This is because of RVP's nature; responders choose to tag a random victim next, which could be the furthest victim away, and this results in more time steps when the number of victims increases. This emphasizes the importance of selecting a quality policy for responders to tag victims adaptively. We also note that although policies LNVP, LCVP, and LGAP assume increasingly uncertain conditions compared to RVP and NVP, they perform well. This suggests that uncertain conditions from inadequate global communication can be addressed by implementing local, iterative, and distributed victim tagging systems for responders, which can lead to more efficient victim tagging times.

Graphs (a) and (b) in Fig. \ref{fig:2graphs} show a policy comparison for the number of victims tagged over each time step for 100 victims and 20 or 80 responders, respectively. It is interesting to see that for both cases, each policy's curve is an S-curve, indicating a sigmoidal relationship between the victims tagged and time. In Fig. \ref{fig:2graphs}(a) NVP, LNVP, and LGAP have steeper curves, which suggests that these policies are more effective in tagging victims quicker as more victims are tagged per each time step. Additionally, the NVP and LNVP curves overlap from the start (\(t = 0\)) until about \(t = 75\), where about 50 victims have been tagged. This indicates that, for any number of victims up to 50, 20 responders tag the same number of victims each time step regardless of whether in a global or local communication scenario. Similarly, other parts of the curves overlap, such as the NVP and LGAP curves at around \(t = 112\), suggesting that policies have similar effectiveness at specific time steps with a particular number of victim and responder agents. 

In both graphs of Fig. \ref{fig:2graphs}, curves overlap, demonstrating that if that particular number of victims and responders exists, then some policies can be interchangeable, thus giving more freedom in the choice of responder policy. In Fig. \ref{fig:2graphs}(b), LGAP performs the least efficiently, taking an average of 148 time steps until all victims are tagged. In LGAP, responders only tag victims within their designated cell, thus some tag more than others. This result highlights the inconsistency of LGAP's performance, as the locations of victims are very influential. Fig. \ref{fig:2graphs}(b) shows that, in comparison to \ref{fig:2graphs}(a), all policies' curves are similar at this scale, indicating that with 80 responders, tagging victims is smoother and more efficient than with 20 responders when there are 100 victims. This indicates that prioritizing policy selection is most important when the ratio of victims to responders is larger. 


Fig. \ref{fig:all} presents results that give additional insights into each policy. Row (I) shows the total victims tagged over each time step, with 5, 20, and 80 responders shown for each policy. All policies have similar curves with slight variations, except RVP has drastic separation between each responder amount, and it has a much larger range of time steps. Comparing the policies, LCVP's curves have a greater disparity at the start of the experiment, as well as a larger difference between the case of 20 and 80 responders. This indicates that LCVP could be less effective at the start of tagging 100 victims, and there is a bigger difference between the ratio of responders to victims for this policy. Results also suggest that if a responder team wants to implement LCVP as a way to prioritize critical victims, then they should strongly consider the number of responders to dispatch in an MCI scenario involving 100 victims.

Row (II) in Fig. \ref{fig:all} shows the average time steps it takes to tag 100 victims for 5, 20, and 80 responders for each policy. NVP results look visually similar to those of LNVP, however for the case with 20 responders the NVP curve increases with a greater slope indicating an inability to tag victims as quickly as LNVP. The ability of responders to communicate locally proves to benefit their iterative process of identifying the next victim to tag.

Row (III) in Fig. \ref{fig:all} shows the average time steps to tag 10, 20, and 100 victims for up to 5 responders for each policy. The graphs look very similar, indicating the common trend of a decrease in the time it takes to tag victims when there are more responders. However, RVP is a much larger scale, taking over 4,000 time steps to tag all victims when there is one responder. Graphs like these can be a valuable tool to determine the number of responders to dispatch based on the number of victims estimated in an MCI. For example, when analyzing all policies except for RVP, the curves converge at similar values. The curves for the cases with 10 and 20 victims look very similar, and at around the point where there are three responders, the lines converge. Therefore, three responders could be a sufficient amount to dispatch for the case where there are either 10 or 20 victims across these policies. This is especially relevant when resources are scarce, and scheduling and resource allocation come into play. Similarly, if there is an existing deadline, such as 1500-time steps, if there are 1-100 victims, it may be sufficient to dispatch 2-5 responders for any of the policies, excluding RVP, based on the graphs. However, if there are more victims or a tighter deadline to accomplish, more responders would be needed to fulfill the time requirement.

Row (IV) in Fig. \ref{fig:all} depicts a responder agent $r_i$'s state for each time step in the case of 5 responders tagging 10 victims for NVP, LNVP, and LGAP (Fig. \ref{fig:fsm}). The NVP graph indicates that each responder was in the tag state for 2 steps, illustrating each responder tagged 2 victims. On the other hand, the local policies, LNVP and LGAP, had varying numbers of tag transitions, demonstrating the adaptation that occurred between responders as they communicated locally and identified victims to tag. The adaptation resulted in improved efficiency for tagging victims quickly as LNVP and LGAP resulted in 118 and 126 time steps to tag all victims, respectively. Results for LGAP further highlight the inconsistency in efficiency due to the locations of victims. Responders 1 and 4 remained idle the entire time, while responders 2 and 3 tagged all the victims.

\begin{table*}[t!]
  \centering
  \normalsize
  \setlength{\tabcolsep}{5pt} 
  \caption{Comparison of FDQN performance with baseline heuristics. The table shows the number of responder agents and victims (No. R, V), with the specific environment size (width $\times$ height) for each experiment. Results highlight the average time steps it takes to tag all victims for each experiment with standard deviation, over fifty iterations. Bold values highlight the most efficient policy for each experiment.}
\begin{tabular}[t]{cc*{7}{c}} 
    \toprule
    \multicolumn{1}{c}{\thead{\small Experiment}} & R1 & R2 & R3 & R4 & R5 & R6 & R7 & R8 \\
        \addlinespace[-2pt] 
        No. R, V & 3, 5 & 3, 10 & 3, 5 & 5, 15 & 5, 50 & 5, 10 & 5, 100 & 20, 100\\
        Env. size & 5 $\times$ 5 & 5 $\times$ 5 & 25 $\times$ 15 & 25 $\times$ 15 & 25 $\times$ 15 & 50 $\times$ 30 & 50 $\times$ 30 & 50 $\times$ 30 \\
        \midrule
      &   &  \multicolumn{5}{c}{\thead{\small Results (average time steps $\pm$ std)}} & \\
    RVP  & 15.7 $\pm$ 1.5 & 26.7 $\pm$ 1.8 & 34.8 $\pm$ 5.2 & 57.4 $\pm$ 6.0 & 159.9 $\pm$ 8.6 & 76.8 $\pm$ 12.0 & 525.9 $\pm$ 23.3 & 160.1 $\pm$ 8.2  \\
    NVP  & 15.4 $\pm$ 1.1 & 24.9 $\pm$ 1.5 & 35.5 $\pm$ 4.7 & \textbf{45.8 $\pm$ 3.5} & 87.3 $\pm$ 4.9 & 69.6 $\pm$ 8.4 & 193.3 $\pm$ 12.6 & 96.4 $\pm$ 6.1  \\
    LNVP & 15.0 $\pm$ 0.8 & 24.4 $\pm$ 1.2 & 34.5 $\pm$ 4.3 & 45.8 $\pm$ 3.4 & \textbf{84.8 $\pm$ 4.7} & 66.2 $\pm$ 6.4 & \textbf{188.3 $\pm$ 12.9} & \textbf{89.0 $\pm$ 3.1}  \\
    LCVP & 15.1 $\pm$ 0.9 & 25.7 $\pm$ 1.6 & 33.8 $\pm$ 5.0 & 49.0 $\pm$ 7.1 & 103.5 $\pm$ 4.3 & \textbf{64.6 $\pm$ 7.4} & 234.9 $\pm$ 14.0 & 114.2 $\pm$ 6.4  \\
    LGAP & 18.6 $\pm$ 3.6 & 28.9 $\pm$ 4.9 & 36.4 $\pm$ 6.8 & 52.0 $\pm$ 7.5 & 97.4 $\pm$ 10.1 & 72.0 $\pm$ 10.0 & 212.4 $\pm$ 16.0 & 107.4 $\pm$ 8.4  \\
    \midrule
    FDQN  & \textbf{12.8 $\pm$ 1.3} & \textbf{20.8 $\pm$ 1.5} & \textbf{33.6 $\pm$ 4.3} & 51.3 $\pm$ 6.5 & 148.4 $\pm$ 8.7 & 70.0 $\pm$ 9.9 & 510.0 $\pm$ 27.3 & 153.7 $\pm$ 8.8  \\
    \bottomrule
\end{tabular}
  \label{tab:results_RL}
\end{table*}

\begin{figure*}[t]
    \centering
    \includegraphics[width=\textwidth]{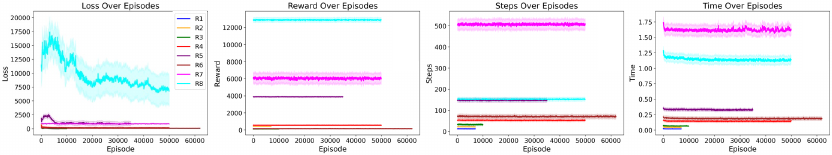}
    \caption{Training graphs for FDQN in experiments R1-R8. The graphs show loss, reward values, number of steps until all victims are tagged, and the computational time in seconds over each training episode. The darker line represents a 71-point simple moving average, smoothing the raw data. The lighter shaded region indicates a $\pm$1 standard deviation from the original (un-smoothed) data.}
    \label{fig:combined_training_graphs}
\end{figure*}

\subsection{Factorized Deep Q-Network Evaluation} \label{sec:FDQN_eval}
DQN‐based approaches can become difficult to manage in large, real‐world environments; hence, prior studies typically focus on smaller grids and fewer agents, rather than the larger, more complex scenarios that reflect true MCIs. To address this gap, we conduct eight experiments spanning a wide range of environment sizes and responder‐to‐victim ratios. These experiments, labeled R1–R8, are summarized in Table \ref{tab:results_RL}.

\subsubsection{Training Details}
For each of these experiments, we implemented a grid search to identify the best-performing hyper-parameters for our training configurations. We optimized for the number of distance bins \(B\in\{5,10\}\), learning rate \(\alpha\in\{0.0005,0.001\}\), discount factor \(\gamma\in\{0.95,0.99\}\), target network update frequency \(f_{\mathrm{update}}\in\{5000,10000\}\), batch size \(\in\{64,128\}\), and decay rate for the exploration parameter \(\epsilon\in\{5000,8000\}\). We additionally utilized a logarithmic schedule for \(\epsilon\) throughout training. The initial exploration rate is set to $\epsilon_{start} = 1.0$, and the final exploration rate is $\epsilon_{final} = 0.1$. We use gradient clipping (max norm = 1.0) to prevent unstable updates and store model checkpoints every 1000 episodes. The optimal hyperparameters found are in Table \ref{tab:rl_hyperparams}. This table also shows the number of episodes each model was trained (until convergence) and the time it took to train each one in minutes. Each model was trained with one NVIDIA A6000 GPU.

\subsubsection{Results}

Table \ref{tab:results_RL} illustrates the performance of our proposed FDQN compared with heuristics. The average number of time steps to tag all victims for fifty iterations of each experiment is listed with standard deviation values.
We include standard deviations here to highlight variability since smaller-scale experiments can yield similar averages, obscuring differences in reliability.
The proposed FDQN outperforms the heuristics in experiments R1-R3, with fewer average time steps needed to tag all victims. This indicates that the multi-agent responder team learned a policy that can tag victims more efficiently than the heuristics.

However, the heuristics outperform the FDQN in R4-R8, which are more complex scenarios. We see that as the ratio of responders to victims and the MCI environment size increases, the FDQN method performs progressively worse in comparison. For example, in R7, where we have our largest ratio (5 responders to 100 victims) in our largest area (50 $\times$ 30), FDQN approaches RVP's performance (selecting random victims to tag).
Comparing R1 to R3, the number of agents is the same, however R3 has a larger environment size of 25 $\times$ 15. FDQN outperforms all methods for both but does comparatively poorer in R3, as it outperforms the next best method (LCVP) by only 0.2 time steps.
This suggests that the FDQN method can be beneficial when there is a smaller ratio of responders to victims and in a smaller area. 

The variances for the FDQN approach increase progressively in a similar way. The standard deviation values for FDQN results increase as the ratio of agents and the environment sizes increase, even surpassing all heuristics' standard deviations for R7 and R8.
While some variance is expected due to the randomness of each experiment run, such as victim locations, a higher variance may indicate instability in the learned policy.

\begin{table}[t]
    \centering
    \small
    \renewcommand{\arraystretch}{1.1}
    \setlength{\tabcolsep}{3pt} 
    \caption{Optimal hyperparameter values found for the FDQN reinforcement learning models in R1-R8, where $B$ is the number of distance bins, $\alpha$ is the learning rate, $\gamma$ is the discount factor, \(f_{\mathrm{update}}\) is the target network update frequency, 'Batch' is the batch size, and '$\epsilon$-decay' is the epsilon decay rate. The number of episodes trained is reported as 'Eps.', which is until convergence. Training time in minutes is listed. One NVIDIA A6000 GPU was used.}
    \footnotesize
    \begin{tabular*}{\columnwidth}{@{\extracolsep{\fill}}lcccccccc} 
        \toprule
         Parameter & R1 & R2 & R3 & R4 & R5 & R6 & R7 & R8 \\
        \midrule
        $B$           & 5   & 10  & 10   & 10   & 5   & 10  & 5   & 10  \\ 
        $\alpha$   & $5e^{-4}$ & $5e^{-4}$& $1e^{-3}$ & $5e^{-4}$ & $5e^{-4}$& $1e^{-3}$  & $1e^{-3}$  & $5e^{-4}$ \\ 
        $\gamma$  & 0.99  & 0.95  & 0.99  & 0.95  & 0.99  & 0.95  & 0.95  & 0.99  \\ 
        \(f_{\mathrm{update}}\)    & 5k  & 5k  & 10k  & 5k  & 5k  & 10k  & 5k  & 10k  \\ 
        Batch       & 64   & 128  & 128   & 128   & 128  & 128  & 64   & 128  \\ 
        $\epsilon$-decay    & 5k  & 5k  & 8k  & 8k  & 5k  & 5k  & 8k  & 8k  \\ 
        \midrule
         Eps.         & 7k  & 7k  & 10k  & 50k  & 35k  & 62k  & 50k  & 50k  \\ 
        Time (min.)      & 2.5   & 5.5  & 11.2   & 120.4   & 193.8  & 192.9  & 1350.1   & 952.9  \\ 
        \bottomrule
    \end{tabular*}
    \label{tab:rl_hyperparams}
\end{table}

\paragraph{Training Insights}
Figure~\ref{fig:combined_training_graphs} compares the FDQN models across all experiments (R1–R8), plotting the training loss, average reward, steps required to tag all victims, and computational time per episode. Each plot applies a 71‐point simple moving average to smooth the raw data, with a $\pm1$ standard deviation band indicating variability.
Overall, the loss curves converge for most runs, but the FDQN in R8 remains significantly higher, likely due to the large environment size of $50 \times 30$, the high agent-to-victim ratio, and the substantial number of responder agents (20 responders and 100 victims). Meanwhile, in R7, FDQN exhibits a lower final loss than in R8 but requires more steps and computational time for each episode, suggesting that ratios of responders to victims can impact training efficiency even more than absolute space size. 
A similar pattern appears when comparing FDQN in R5 (in a smaller $25 \times 15$ grid) to R6 (in a larger $50 \times 30$ grid): despite a smaller physical area, FDQN in R5 takes more steps to finish tagging and longer to train, again pointing to the importance of agent to victim ratios on FDQN training outcomes. 

The reward structure is the same for all FDQN experiments, so the values vary per experiment solely influenced by the number of agents and environment size, similar to other metrics. The computational time per experiment highlights the same pattern, indicating that internal computations become more expensive as the agent ratio and space grows. Figure \ref{fig:all_fdqn} in the Appendix depicts training graphs detailing each FDQN experiment to show underlying patterns in the data.



\section{Discussion and Conclusion} \label{conc}
This paper aims to address the multi-agent victim tagging problem of minimizing the time it takes to tag all victims during an MCI. We formalized the victim tagging problem using ILP,
and proposed five applicable, distributed, on-the-go heuristics considering local and global communication constraints. We further designed and implemented a factorized deep Q-network approach to learn an optimal responder team strategy and compared performance with the baseline heuristics. Our solutions were evaluated through a series of simulation experiments for between-heuristic comparisons and extended simulation experiments for comparing with the FDQN approach. 

\subsection{Heuristics Findings}
The between-heuristic experiments gave us insights into rule-based approaches and the impact of local vs. global communication constraints. Results demonstrated that the local policies performed most efficiently in identifying the next victim to tag for each responder in an on-the-go fashion. Specifically, LNVP consistently performed most efficiently, and RVP performed least efficiently. 

Individual policy analyses provided further insights. For policies assuming global communication, NVP performed significantly better than RVP. The employment of local adaptive tagging proved valuable for selecting the next victim to tag. LCVP performed well but not the most efficiently. If the goal was instead to prioritize critical victims with a performance metric involving lives saved, LCVP could be explored as a potential optimal policy. LGAP performed poorly compared to the rest of the policies, as its performance depended on the locations of victims. An extension of LGAP where the responder cells consider victim locations could be an improvement to consider in future work. 


Our heuristics‐based results can serve as practical guidelines for a range of victim tagging setups, including fully autonomous responder teams or mixed human-robot collaborations. In situations with limited communication, these heuristics remain robust by offering predefined rules that minimize coordination overhead. Because emergency departments typically rely on ad hoc methods rather than explicit victim tagging algorithms, responders who know or estimate the number of victims can reference these results to select an appropriate victim tagging policy and also decide how many responders to deploy. This approach helps ensure a more consistent, time‐sensitive response in MCIs.


\subsection{Factorized Deep Q-Network Findings}
Our FDQN method outperformed heuristics in smaller‐scale MCIs and shows promise for scaling to larger environments and higher agent counts. Unlike many traditional RL studies, we explored upper limits by running eight diverse scenarios, some with expansive areas and sizeable agent teams, thus providing insights into the performance ceiling of this approach. Such exploration benefits two primary audiences: (1) RL researchers extending factorized multi‐agent solutions to increasingly complex domains, and (2) emergency management practitioners aiming to integrate autonomous decision support in critical, real‐time response.

Our FDQN method demonstrates that multi‐agent factorization, where each responder selects sub‐actions individually but shares a global state, can keep action spaces tractable in large, realistic environments. However, certain design decisions reveal potential bottlenecks. For instance, a fixed 5- or 10-bin discretization in larger environments loses fine-grained distance information. This likely contributed to more unstable learning and weaker policies in runs with large spaces. Similarly, higher responder-to-victim ratios may reduce effective exploration by leading to redundant sub-action choices (e.g., selecting the same victim).
Future work could include adaptive binning, improved inter-agent communication, and advanced knowledge sharing to foster earlier coordination and deeper cooperation.
Moreover, our factorization approach assumes sub-action (actions for each agent) independence, which may not hold when sub-actions interact significantly; this can introduce bias in Q-function estimates, suggesting a need for explicit interaction terms or more complex methods in such cases.
Future directions include improved inter‐agent communication protocols, deeper cooperation strategies, and further analysis of large‐scale viability.


For emergency response management, this study underscores the feasibility of integrating MARL solutions into crisis response scenarios. Our experiments suggest that learned policies can enhance or supplement human decision‐making, particularly when multiple responders coordinate to tag victims efficiently. While the FDQN approach alone does not solve all collaboration challenges, it provides an important step toward autonomous systems capable of adaptive, goal‐oriented behavior in real‐time emergencies. The interplay between decentralized action selection and globally informed training underscores the potential for robust, scalable solutions applicable to diverse disaster response domains.


{\appendix[Training Graphs]
See detailed training graphs below for each FDQN experiment R1-R8.


\begin{figure*}[t]
    \centering
    \includegraphics[width=0.85\textwidth]{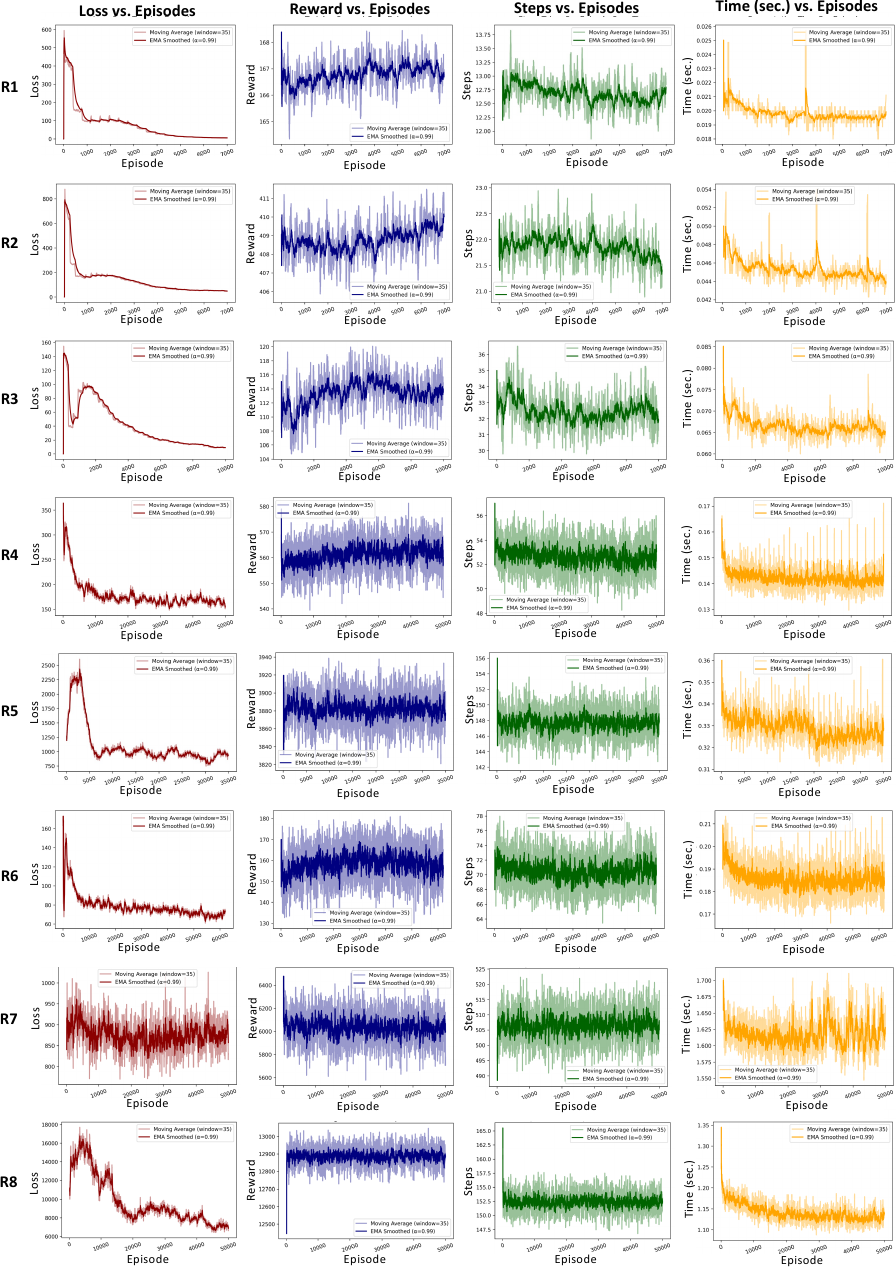}
    \caption{Detailed training graphs for each experiment R1-R8. Each column reports different metrics over each episode. The darker line is the exponential moving average with a smoothing factor of 0.99, and the lighter line represents the simple moving average over a window size of 35. These smoothing techniques enable detailed trend analysis by reducing noise and highlighting underlying patterns in the data.}
    \label{fig:all_fdqn}
\end{figure*}

}


 
%

\bibliographystyle{IEEEtran}

\bibliography{my_paper}

\begin{thebibliography}{10}
\providecommand{\url}[1]{#1}
\csname url@samestyle\endcsname
\providecommand{\newblock}{\relax}
\providecommand{\bibinfo}[2]{#2}
\providecommand{\BIBentrySTDinterwordspacing}{\spaceskip=0pt\relax}
\providecommand{\BIBentryALTinterwordstretchfactor}{4}
\providecommand{\BIBentryALTinterwordspacing}{\spaceskip=\fontdimen2\font plus
\BIBentryALTinterwordstretchfactor\fontdimen3\font minus \fontdimen4\font\relax}
\providecommand{\BIBforeignlanguage}[2]{{%
\expandafter\ifx\csname l@#1\endcsname\relax
\typeout{** WARNING: IEEEtran.bst: No hyphenation pattern has been}%
\typeout{** loaded for the language `#1'. Using the pattern for}%
\typeout{** the default language instead.}%
\else
\language=\csname l@#1\endcsname
\fi
#2}}
\providecommand{\BIBdecl}{\relax}
\BIBdecl

\bibitem{ems_mci}
\BIBentryALTinterwordspacing
R.~L. DeNolf and C.~I. Kahwaji, ``Statpearls,'' Jan 2023, a service of the National Library of Medicine, National Institutes of Health. [Online]. Available: \url{https://www.ncbi.nlm.nih.gov/books/NBK482373/}
\BIBentrySTDinterwordspacing

\bibitem{Hawe}
G.~I. Hawe, G.~Coates, D.~T. Wilson, and R.~S. Crouch, ``Agent-based simulation of emergency response to plan the allocation of resources for a hypothetical two-site major incident,'' in \emph{Innovative Artificial Intelligence Solutions for Crisis Management}, vol.~46, 2015, pp. 336--345.

\bibitem{Gonzalez}
R.~A. Gonzalez, ``A framework for ict-supported coordination in crisis response,'' Ph.D. dissertation, Delft University of Technology, Delft, Netherlands, 2010.

\bibitem{Gonzalez-2}
------, ``Analysis and design of a multi-agent system for simulating a crisis response organization,'' in \emph{Enterprises \& Organizational Modeling and Simulation}, no.~6, 2009, pp. 1--15.

\bibitem{Wang}
Y.~Wang, L.~K. Luangkesorn, and L.~Shuman, ``Modeling emergency medical response to a mass casualty incident using agent-based simulation,'' \emph{Socio-Economic Planning Sciences}, vol.~46, no.~4, pp. 281--290, 2012.

\bibitem{Bellamine}
N.~Bellamine Ben~Saoud, T.~Mena, J.~Dugdale, B.~Pavard, and M.~Ben~Ahmed, ``Assessing large scale emergency rescue plans: an agent based approach,'' \emph{Int. J. Intell. Control Syst}, vol.~11, 12 2006.

\bibitem{tomczyk2023}
L.~Tomczyk and Z.~Kulesza, ``Multiple-criteria decision-making for medical rescue operations during mass casualty incidents,'' \emph{Applied Sciences}, vol.~13, no.~13, p. 7467, 2023.

\bibitem{lee2021}
H.-R. Lee and T.~Lee, ``Multi-agent reinforcement learning algorithm to solve a partially-observable multi-agent problem in disaster response,'' \emph{European Journal of Operational Research}, vol. 291, no.~1, pp. 296--308, 2021.

\bibitem{zhou2023}
W.~Zhou, C.~Zhang, and S.~Chen, ``Dual deep q-learning network guiding a multiagent path planning approach for virtual fire emergency scenarios,'' \emph{Applied Intelligence}, vol.~53, pp. 21\,858--21\,874, 2023.

\bibitem{yang2020}
Z.~Yang, L.~Nguyen, J.~Zhu, Z.~Pan, J.~Li, and F.~Jin, ``Coordinating disaster emergency response with heuristic reinforcement learning,'' in \emph{2020 IEEE/ACM International Conference on Advances in Social Networks Analysis and Mining (ASONAM)}, 2020, pp. 565--572.

\bibitem{CASE_paper}
M.~A. Cardei and A.~Doryab, ``Practical heuristics for victim tagging during a mass casualty incident emergency medical response,'' in \emph{2024 IEEE 20th International Conference on Automation Science and Engineering (CASE)}, 2024, pp. 165--172.

\bibitem{Turner}
J.~Turner, Q.~Meng, and G.~Schaefer, ``Increasing allocated tasks with a time minimization algorithm for a search and rescue scenario,'' in \emph{2015 IEEE International Conference on Robotics and Automation (ICRA)}, 2015, pp. 3401--3407.

\bibitem{Geng}
N.~Geng, Q.~Meng, D.~Gong, and P.~W.~H. Chung, ``How good are distributed allocation algorithms for solving urban search and rescue problems? a comparative study with centralized algorithms,'' \emph{IEEE Transactions on Automation Science and Engineering}, vol.~16, no.~1, pp. 478--485, 2019.

\bibitem{zhao}
W.~Zhao, Q.~Meng, and P.~W.~H. Chung, ``A heuristic distributed task allocation method for multivehicle multitask problems and its application to search and rescue scenario,'' \emph{IEEE Transactions on Cybernetics}, vol.~46, no.~4, pp. 902--915, 2016.

\bibitem{Shi}
\BIBentryALTinterwordspacing
J.~Shi, Z.~Yang, and J.~Zhu, ``An auction-based rescue task allocation approach for heterogeneous multi-robot system,'' \emph{Multimedia Tools Appl.}, vol.~79, no. 21–22, p. 14529–14538, Jun 2020. [Online]. Available: \url{https://doi.org/10.1007/s11042-018-7080-4}
\BIBentrySTDinterwordspacing

\bibitem{Chen}
J.~Chen, C.~Du, X.~Lu, and K.~Chen, ``Multi-region coverage path planning for heterogeneous unmanned aerial vehicles systems,'' in \emph{2019 IEEE International Conference on Service-Oriented System Engineering (SOSE)}, 2019, pp. 356--3565.

\bibitem{choi2009}
H.-L. Choi, L.~Brunet, and J.~P. How, ``Consensus-based decentralized auctions for robust task allocation,'' \emph{IEEE Transactions on Robotics}, vol.~25, no.~4, pp. 912--926, 2009.

\bibitem{Luo}
C.~Luo, Q.~Huang, F.~Kong, S.~Khan, and Q.~Qiu, ``Applying machine learning in designing distributed auction for multi-agent task allocation with budget constraints,'' in \emph{2021 20th International Conference on Advanced Robotics (ICAR)}, 2021, pp. 356--363.

\bibitem{Bramblett}
L.~Bramblett, R.~Peddi, and N.~Bezzo, ``Coordinated multi-agent exploration, rendezvous, \& task allocation in unknown environments with limited connectivity,'' in \emph{2022 IEEE/RSJ International Conference on Intelligent Robots and Systems (IROS)}, 2022, pp. 12\,706--12\,712.

\bibitem{Pallin}
M.~Pallin, J.~Rashid, and P.~Ögren, ``A decentralized asynchronous collaborative genetic algorithm for heterogeneous multi-agent search and rescue problems,'' in \emph{2021 IEEE International Symposium on Safety, Security, and Rescue Robotics (SSRR)}, 2021, pp. 1--8.

\bibitem{sutton2018}
R.~S. Sutton and A.~G. Barto, \emph{Reinforcement Learning: An Introduction}, 2nd~ed.\hskip 1em plus 0.5em minus 0.4em\relax MIT Press, 2018.

\bibitem{watkins1992}
C.~J. C.~H. Watkins and P.~Dayan, ``Q-learning,'' \emph{Machine Learning}, vol.~8, no. 3-4, pp. 279--292, 1992.

\bibitem{mnih2015}
V.~Mnih, K.~Kavukcuoglu, D.~Silver, A.~A. Rusu, J.~Veness, M.~G. Bellemare, A.~Graves, M.~Riedmiller, A.~K. Fidjeland, G.~Ostrovski, S.~Petersen, C.~Beattie, A.~Sadik, I.~Antonoglou, H.~King, D.~Kumaran, D.~Wierstra, S.~Legg, and D.~Hassabis, ``Human-level control through deep reinforcement learning,'' \emph{Nature}, vol. 518, pp. 529--533, 2015.

\bibitem{oroojlooyjadid2019}
\BIBentryALTinterwordspacing
A.~OroojlooyJadid and D.~Hajinezhad, ``A review of cooperative multi-agent deep reinforcement learning,'' \emph{arXiv preprint arXiv:1908.03963}, 2019. [Online]. Available: \url{https://arxiv.org/abs/1908.03963}
\BIBentrySTDinterwordspacing

\bibitem{azadeh2024}
\BIBentryALTinterwordspacing
R.~Azadeh, ``Advances in multi-agent reinforcement learning: Persistent autonomy and robot learning lab report 2024,'' 2024. [Online]. Available: \url{https://arxiv.org/abs/2412.21088}
\BIBentrySTDinterwordspacing

\bibitem{Hernandez-Leal2019}
\BIBentryALTinterwordspacing
P.~Hernandez-Leal, B.~Kartal, and M.~E. Taylor, ``A survey and critique of multiagent deep reinforcement learning,'' \emph{Autonomous Agents and Multi-Agent Systems}, vol.~33, no.~6, p. 750–797, Nov. 2019. [Online]. Available: \url{https://doi.org/10.1007/s10458-019-09421-1}
\BIBentrySTDinterwordspacing

\bibitem{nguyen2020}
T.~T. Nguyen, N.~D. Nguyen, and S.~Nahavandi, ``Deep reinforcement learning for multiagent systems: A review of challenges, solutions, and applications,'' \emph{IEEE Transactions on Cybernetics}, vol.~50, no.~9, pp. 3826--3839, 2020.

\bibitem{Sunehag2017}
\BIBentryALTinterwordspacing
P.~Sunehag, G.~Lever, A.~Gruslys, W.~M. Czarnecki, V.~F. Zambaldi, M.~Jaderberg, M.~Lanctot, N.~Sonnerat, J.~Z. Leibo, K.~Tuyls, and T.~Graepel, ``Value-decomposition networks for cooperative multi-agent learning,'' \emph{ArXiv}, vol. abs/1706.05296, 2017. [Online]. Available: \url{https://api.semanticscholar.org/CorpusID:25026734}
\BIBentrySTDinterwordspacing

\bibitem{rashid2020}
T.~Rashid, M.~Samvelyan, C.~S. De~Witt, G.~Farquhar, J.~Foerster, and S.~Whiteson, ``Monotonic value function factorisation for deep multi-agent reinforcement learning,'' \emph{J. Mach. Learn. Res.}, vol.~21, no.~1, Jan. 2020.

\bibitem{son2019}
\BIBentryALTinterwordspacing
K.~Son, D.~Kim, W.~J. Kang, D.~E. Hostallero, and Y.~Yi, ``Qtran: Learning to factorize with transformation for cooperative multi-agent reinforcement learning,'' 2019. [Online]. Available: \url{https://arxiv.org/abs/1905.05408}
\BIBentrySTDinterwordspacing

\bibitem{yang2020Qatten}
\BIBentryALTinterwordspacing
Y.~Yang, J.~Hao, B.~Liao, K.~Shao, G.~Chen, W.~Liu, and H.~Tang, ``Qatten: A general framework for cooperative multiagent reinforcement learning,'' 2020. [Online]. Available: \url{https://arxiv.org/abs/2002.03939}
\BIBentrySTDinterwordspacing

\bibitem{MTZ}
\BIBentryALTinterwordspacing
G.~Pataki, ``Teaching integer programming formulations using the traveling salesman problem,'' \emph{SIAM Rev.}, vol.~45, pp. 116--123, 2003. [Online]. Available: \url{https://api.semanticscholar.org/CorpusID:15997192}
\BIBentrySTDinterwordspacing

\bibitem{mtz2}
\BIBentryALTinterwordspacing
C.~E. Miller, A.~W. Tucker, and R.~A. Zemlin, ``Integer programming formulation of traveling salesman problems,'' \emph{J. ACM}, vol.~7, no.~4, p. 326–329, oct 1960. [Online]. Available: \url{https://doi.org/10.1145/321043.321046}
\BIBentrySTDinterwordspacing

\bibitem{START}
\BIBentryALTinterwordspacing
L.~A.~F. Department, ``Start - simple triage and rapid treatment,'' 2005, accessed: 2024-02-28. [Online]. Available: \url{https://www.cert-la.com/downloads/education/english/start.pdf}
\BIBentrySTDinterwordspacing

\bibitem{mesa}
J.~Kazil, D.~Masad, and A.~Crooks, ``Utilizing python for agent-based modeling: The mesa framework,'' in \emph{Social, Cultural, and Behavioral Modeling}, R.~Thomson, H.~Bisgin, C.~Dancy, A.~Hyder, and M.~Hussain, Eds.\hskip 1em plus 0.5em minus 0.4em\relax Cham: Springer International Publishing, 2020, pp. 308--317.

\bibitem{REMSC}
\BIBentryALTinterwordspacing
R.~E. Council, ``Regional mass casualty incident plan,'' accessed: 2024-02-28. [Online]. Available: \url{https://remscouncil.org/wp-content/uploads/2022/03/REMS-Regional-MCI-Plan-February-2022.pdf}
\BIBentrySTDinterwordspacing

\end{thebibliography}

\vfill

\end{document}